\documentclass[twocolumn,floatfix,showpacs,amsmath,superscriptaddress]{revtex4}
\usepackage{graphicx}
\usepackage{bm}
\usepackage{epsf}
\DeclareGraphicsExtensions{.eps}
\begin{document}
\title{Dispersion and fidelity in quantum interferometry}
\author{D.S. Simon}
\affiliation{Dept. of Electrical and Computer Engineering, Boston University, 8 St. Mary's St., Boston, MA 02215}
\author{A.V. Sergienko}
\affiliation{Dept. of Electrical and Computer Engineering, Boston University, 8 St. Mary's St., Boston, MA 02215}
\affiliation{Dept. of Physics, Boston University, 590 Commonwealth Ave., Boston, MA 02215}
\author{T.B. Bahder}
\affiliation{Charles M. Bowden Research Facility, US Army RDECOM, Aviation and Missile Research, Development,
and Engineering Center, Redstone Arsenal, AL 35898}

\date{\today}

\begin{abstract}
We consider Mach-Zehnder and Hong-Ou-Mandel interferometers with nonclassical states of light as input, and study the
effect that dispersion inside the interferometer has on the sensitivity of phase measurements. We study in detail a number of different one- and
two-photon input states, including Fock, dual Fock, N00N states, and photon pairs from parametric downconversion.
Assuming there is a phase shift $\phi_0$ in one arm of the interferometer, we compute the probabilities of
measurement outcomes as a function of $\phi_0$, and then compute the Shannon mutual information between
$\phi_0$ and the measurements. This provides a means of quantitatively comparing the utility of various
input states for determining the phase in the presence of dispersion. In addition, we consider a simplified model of parametric
downconversion for which probabilities can be explicitly computed analytically, and which serves as a limiting case of
the more realistic downconversion model.
\end{abstract}


\maketitle

\section{Introduction} Interferometry is both an important tool for practical
measurements and a useful testing ground for fundamental physical principles. As a result,
the search for methods to improve the resolution of interferometers forms an active area of
study. It has been shown by a number of
authors (\cite{yurke1, yurke2, holland, boll})
that nonclassical states, in particular those with high degrees of
entanglement, when used as input to an interferometer can lead to resolutions that approach the Heisenberg
limit, the fundamental physical limit imposed by the uncertainty principle.
Most of this previous work has dealt with idealized interferometers,
with no dispersion or photon losses. Before quantum interferometry may become a
useful practical tool the question must be asked as to how well the conclusions
of these previous studies hold up in more realistic and less idealized situations. In this paper, we will attempt to take
the next step along this road by adding dispersion to the apparatus and examining what effect this has on the phase sensitivity
of interferometry with nonclassical input. The motivation for this work is the desire
to ultimately construct quantum sensors that can measure the values of external fields
by measuring the phases shifts they produce in an interferometer.

In particular, the nonclassical input states we will consider are (i) Fock states $|N,0\rangle$ which have a fixed number of photons incident
on one input port, (ii) dual or twin Fock states $|N,N\rangle$ which have the same
number of photons incident on each input port, and (iii) N00N states
$ {1\over \sqrt{2}}\left[ |N,0\rangle +|0,N\rangle\right] $. Here, $|N_a,N_b\rangle $ denotes a state with numbers $N_a$ and
$N_b$ of photons entering each of the two interferometer input ports.

There has been a great deal of recent work on the production of
nonclassical states of light with large ($N>2$) numbers of photons
by means of postselection (for example, \cite{knill, dowl,
walther, mitchell}), however at present the utility of these
postselection schemes for application to practical situations is
not clear. Although this work is useful for clarifying the
scientific issues involved, it is not technologically feasible at
present to use these methods to produce the desired states on
demand. Rather, postselection produces states statistically, at
random times, and therefore can not be relied upon to produce
states on demand for a quantum sensor. In addition, for large
photon number, great care must be taken to distinguish between
states of $N$ photons and those of $N-1$ photons, making it
difficult to prevent mixed states from appearing, which would
change the physics involved. In contrast, two-photon entangled
states with well-defined properties can be easily produced by
parametric downconversion or other methods.

Due to the current practical difficulties of producing on demand entangled photon states with large, well-defined
$N$, we save the large $N$ case for later study and
restrict ourselves in this current paper to situations which are both simpler and
of more immediate practical interest, namely the cases of one or two
photons. Furthermore, for the two-photon case, we consider two possibilities: (i) the photons may
be uncorrelated in frequency, or (ii)
the pair may be produced through spontaneous parametric downconversion (SPDC), resulting
in anticorrelation between the two frequencies.

Our goal is to compare the usefulness of each of these cases for making phase measurements in the presence of
dispersion, so we will need a means of quantifying the sensitivity of the interferometer
with respect to these measurements. Consider a single shot consisting of a nonclassical state of light with
a fixed number of photons being injected into the input ports of the interferometer.
Suppose some phase-dependent observable $M(\phi )$ is measured during this shot.
The usual way to define the phase sensitivity
of the measurement is by computing
\begin{equation}\Delta\phi =\left| {{dM}\over {d\phi}}
\right|^{-1}\Delta M.\end{equation} However, this is correct only
if the probability distribution of the phases has a single peak
and is approximately Gaussian in shape. A more general strategy is
to take an information-theoretical approach and to define the
quantum fidelity by means of the Shannon mutual information
\cite{bahdlap} {\small \begin{equation}H(\Phi :M)={1\over {2\pi}}
\sum_m \int_{-\pi}^\pi d\phi \; P(m|\phi ) \log_2\left[ {{2\pi
P(m|\phi )}\over {\int_{-\pi}^\pi  P(m|\phi^\prime )d\phi^\prime}}
\right] . \label{info}\end{equation}} Here, $m$ and $\phi$ are the
measured values of the random variables $M$ and $\Phi$, while
$P(m|\phi )$ is the conditional probability of obtaining
measurement $m$ given the phase $\phi$ on a particular shot.
In this formula, we have also assumed maximum ignorance of the phase,
i.e., we have assumed a uniform distribution for $\phi$,
$p(\phi )={1\over {2\pi}}$.
Suppose that the detectors have a characteristic time-scale $T_D$.
Then in this context, a single shot will consist of a well-defined
number of photons entering the apparatus simultaneously (i.e.
within a temporal window much smaller than $T_D$) and seperated in
time from any other entering photons by a time $\ge T_D$. The
mutual information is a measure of the information gained per shot
about the phase $\Phi$ from a measurement of the observable $M$.
In our case, the role of $M$ will be played by the number of
photons detected at each of the output ports. For $N$ input
photons, output detector C will count $l$ photons, detector $D$
will detect the remaining $N-l$ photons, and the sum in equation
\ref{info} will become a sum over $l$, where $l=0,1,\dots N$.
Throughout this paper we will use the quantum fidelity as our
measure of phase sensitivity. Besides being of very general
applicability and giving a precise, calculable measure for the
utility of a measurement, the introduction of the mutual
information provides a link to the theory of quantum information
processing. Bahder and Lopata \cite{bahdlap} have computed the
quantum fidelity as a function of $N$ for idealized lossless and
dispersionless interferometers with Fock and N00N state inputs. In
the following sections, we will see how their results change for
the cases of $N=1$ and $N=2$ when dispersion is present.

Although not the principal focus of this paper, it should be
noted that the existence of multiple
peaks in the output probability distributions invalidate the assumptions used to
derive the Heisenberg bound from the Cramer-Rao lower bound, which makes input
states with multimodal distributions especially interesting from the point
of view of the study of phase sensitivity. Note that violations of the
Heisenberg limit have recently been shown to exist in another context, distinct from the
situation examined in this paper, namely in the context of nonlinear
interferometry \cite{luis, beltran, boixo}.

We will assume one branch of the interferometer has a dispersive element
which gives the photon wavenumber $k$ a frequency dependence of the form
\begin{equation}k(\omega )=k_0+\alpha (\omega -\omega_0) +\beta (\omega
-\omega_0)^2 ,\label{dispersiondef}\end{equation}
ignoring the possibility of higher order terms. The other interferometer
arm will be assumed to be of negligible dispersion. Here, $\alpha$ is the inverse of
the group velocity, and $\beta$ is the group delay dispersion per unit length.

In addition to the Mach-Zehnder interferometer, we will examine the fidelity of an alternate setup used in \cite{dowl},
in which N00N states are incident on a single beam-splitter used as a Hong-Ou-Mandel (HOM) interferometer. We will then be in a
position to compare the possible input states and interferometer setups, with a view to gaining insight into their
relative usefulness in practical measurements. In the two-photon cases, we must distinguish
between situations in which the photon energies (or frequencies) are correlated and those in which they are independent.
Thus, after we examine the case of energy-uncorrelated photons, we look at photon pairs anticorrelated in
energy. We further consider two subcases of the latter: (i) a simple model which can be solved
analytically and which amounts to a simplified version of spontaneous downconversion, and (ii)
a more realistic but less analytically tractable version of downconversion.

The plan of this paper is as follows:
in section \ref{mz} we consider the setup for the dispersive Mach-Zehnder
interferometer and define the input states we will use in more detail.
In section \ref{mz1}, we apply the possible one-photon
inputs to the interferometer and compute the probabilities
for the various possible outcomes. In sections \ref{mz2} and
\ref{hom} respectively, we do the same for the Mach-Zehnder interferometer
with several different two-photon inputs and for
the HOM interferometer with $N=2$ N00N state input. In section \ref{comp} we compute and plot the mutual information for
each of the preceeding cases as functions of bandwidth and dispersion levels; we then compare and discuss the results for
the various cases. Finally, in section \ref{spdcreal} we repeat the same calculation for input consisting of a photon
pair produced via spontaneous parametric downconversion before arriving at final conclusions in section \ref{conc}.

For ease of reference later, table I summarizes the specific cases we will examine over the following sections.
\begin{table}
\caption{\label{tab:table1} Summary of the special cases examined in the later sections of this paper.}
\begin{ruledtabular}
\begin{tabular}{|c|c|c|c|c|}\hline
Case & \# of   & Interferometer  &  Input & Frequency \\
No.   & photons &  Type           & State & Correlation \\ \hline \hline
A     & 1 &  MZ & Fock & not applicable \\ \hline
B    & 1 &  MZ &  N00N & not applicable \\  \hline
C   & 2 &  MZ & Fock & none \\ \hline
D  & 2 &  MZ & Dual Fock & none \\ \hline
E   & 2 &  MZ & N00N &  none \\ \hline
F  & 2 &  MZ & Fock & anticorrelated \\ \hline
G  & 2 &  MZ & N00N &  anticorrelated \\ \hline
H & 2 &  MZ & Dual Fock & anticorrelated \\ \hline
I   & 1 & HOM & N00N & none \\ \hline
J    & 2 & HOM & N00N & none \\ \hline
K   & 2 & HOM & N00N & anticorrelated \\ \hline
L  & 2 &  MZ & SPDC Fock & anticorrelated \\ \hline
\end{tabular}
\end{ruledtabular}
\end{table}

\section{The Dispersive Mach-Zehnder Interferometer}\label{mz}

\begin{figure}
\centering
\includegraphics[totalheight=2in]{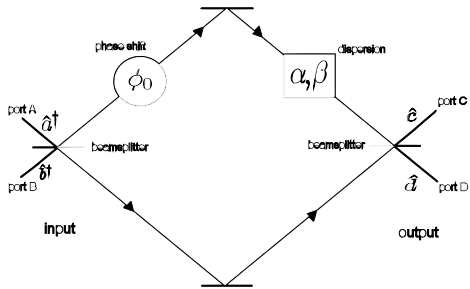}
\caption{Mach-Zehnder interferometer with dispersion in one arm. There is
also a phase shift $\phi_0$ of nondispersive origin in the same arm.}
\label{mzfig}
\end{figure}

Consider the Mach-Zehnder interferometer of figure \ref{mzfig}, with 50/50 beamsplitters.
Assume for the moment that there is no dispersion in the apparatus.
Let $\hat a_\omega$ and $\hat b_\omega$ be operators that annihilate photon states in the
two input ports A and B. They obey the usual canonical commutation
relations with the corresponding creation operators
$\hat a_\omega^\dagger$ and $\hat b_\omega^\dagger$:
\begin{equation} \left[ \hat a_\omega,\hat a_{\omega^\prime}^\dagger\right]
=\left[ \hat b_\omega,\hat b_{\omega^\prime}^\dagger\right] =\delta (\omega -\omega^\prime )
,\end{equation} with all other commutators vanishing. For independent photons,
the input states to the interferometer can be described
in terms of the number of photons entering the two ports: \begin{eqnarray}& & | N_a,N_b; \omega_1,\dots ,
\omega_{N_a};\omega^\prime_1\dots ,\omega^\prime_{N_b}\rangle \nonumber\\
& & \qquad ={1\over \sqrt{N_a! N_b!}}\hat a^\dagger_{\omega_1}\dots \hat
a^\dagger_{\omega_{N_a}}
\hat b^\dagger_{\omega^\prime_1}\dots \hat b^\dagger_{\omega^\prime_{N_b}}
|0\rangle ,\end{eqnarray} where $N_a$ and $N_b$ are the number of photons
in ports A and B, respectively, and $|0\rangle$ is the vacuum
state with no photons. Similarly, $N_c$, $N_d$, $\hat c_\omega$, and $\hat d_\omega$ will represent
the photon numbers and annihilation operators at output ports C and D.

The effect of the Mach-Zehnder interferometer on a given input state may be described in terms of the scattering
matrix, $S(\phi )$.
The initial and final annihilation operators
are related by a scattering matrix $S(\phi)$:
\begin{equation}\left( \begin{array}{c}\hat c_\omega (\phi) \\ \hat d_\omega (\phi)\end{array}\right) =S(\phi )
\left( \begin{array}{c}\hat a_\omega \\ \hat b_\omega \end{array}\right) ,\label{scatmatrix}\end{equation}
where $\phi$ is the
relative phase difference experienced by photons in the two arms.
In the absence of photon losses in the system, the scattering matrix will be
unitary. Then,
for an ideal Mach-Zehnder interferometer, the scattering matrix is given by
\begin{eqnarray}S(\phi ) &=& {1\over 2}\left[ e^{i\phi} e^{ikL_1} -e^{ikL_2}\right] \sigma_z
-{i\over 2} \left[ e^{i\phi} e^{ikL_1} + e^{ikL_2}\right] \sigma_x\nonumber \\
&=& -i e^{ikL_1}e^{i\phi /2}\left(\begin{array}{cc}-\sin{\phi\over 2}
& \cos {\phi\over 2} \\ \cos{\phi\over 2} & \sin{\phi\over 2}
\end{array}\right) , \end{eqnarray} where the Pauli matrices are \begin{equation} \sigma_x=
\left( \begin{array}{cc} 0 & 1\\ 1 & 0\\ \end{array} \right) \qquad
\mbox{and}\qquad  \sigma_z=
\left( \begin{array}{cc} 1 & 0\\ 0 & -1\\ \end{array} \right) .
\end{equation} In this scattering matrix we have assumed (as we will assume henceforth)
that the lengths of the two interferometer arms are equal, $
L_1=L_2$. Using this matrix in equation~\ref{scatmatrix}, we can invert the equation and
take adjoints to arrive at the following result:
\begin{eqnarray}\hat a_\omega^\dagger &=& i\left[ \hat c_\omega^\dagger \sin {\phi\over 2} -
\hat d_\omega^\dagger \cos {\phi\over 2} \right] e^{i\phi /2} \label{adef}\\
\hat b_\omega^\dagger &=& -i\left[ \hat c_\omega^\dagger \cos
{\phi \over 2} + \hat d_\omega^\dagger \sin {\phi \over 2} \right]
e^{i\phi /2} \label{bdef}\end{eqnarray}

We assume that the frequency distribution for each incoming photon
is Gaussian and that each Gaussian has the same width and central
frequency, of the form $e^{- {1\over 2}\sigma (\omega
-\omega_0)^2}$. Input and output states will either be states of
definite photon number in the sense that they are eigenstates of
number operators of the form $\hat N_j=\int d\omega
a^{(j)\dagger}_\omega a^{(j)}_\omega$ (where $a_\omega^{(j)}$ is
the annihilation operator for photons at the {\it j}th port), or
else superpositions of such states.

We introduce dispersion to the upper branch of the interferometer
by giving the wavenumber $k$ a frequency dependence of the form in
equation \ref{dispersiondef}. We assume that the dispersion in the
other branch of the interferometer is negligible, i.e. that
$k(\omega )=k_0$ in that branch. The length of the portion of the
upper arm for which $k(\omega)$ differs from $k_0$ will be denoted
$L$, where $0\le L\le L_1$. In addition to any phase difference
resulting from the asymmetric dispersion, we also assume that
photons travelling through the upper branch of the interferometer
gain an additional phase difference $\phi_0$ relative to the lower
branch. $\phi_0$ is any phase difference of nondispersive origin
that may be present in the setup; this may be due to a difference
in path length, or an interaction of one arm of the interferometer
with an external field. Note that for our setup, the assumption of
a balanced interferometer entails no loss of generality; to
account for an unbalanced interferometer, it suffices to simply
include a term of the form $k_0(L_1-L_2)$ inside the phase factor
$\phi_0$.

In the presence of the dispersion, the scattering matrix will now be of the form \begin{eqnarray}
S(\phi_0) &=&  {1\over 2} e^{ik_0L_1}\left( \begin{array}{cc}
e^{i\phi (\omega  )}-1 & -i\left( e^{i\phi (\omega  )}+1\right)\\
-i\left( e^{i\phi (\omega  )}+1\right) &
-\left( e^{i\phi (\omega  )}-1\right) \end{array}\right) \nonumber\\
&=& -ie^{ik_0L_1}e^{i\phi (\omega )/2} \left( \begin{array}{cc}
-\sin{{\phi (\omega )}\over 2} & \cos{{\phi (\omega )}\over 2} \\
\cos{{\phi (\omega )}\over 2} & \sin{{\phi (\omega )}\over 2}
\end{array}\right)  \end{eqnarray} where for future convenience
we have shifted the frequency dependence into a new phase angle by defining \
\begin{equation}\phi (\omega )=\phi_0 +\alpha L(\omega -\omega_0)
+\beta L (\omega -\omega_0 )^2.\label{phiomega}\end{equation}

Consider $N$ photons entering the interferometer and assume for now that their frequencies are independent variables.
The Fock, dual Fock, and N00N input states are of the form:\begin{eqnarray}|N,0\rangle_\sigma
&=& {1\over \sqrt{N!}}\left( {{\sigma}\over \pi}\right)^{N/4} \int d\omega_1\dots d\omega_N e^{- {\sigma\over 2}
\sum_{j=1}^N
(\omega_j -\omega_0)^2}\nonumber\\
& & \qquad \qquad \qquad \times \hat a^\dagger_{\omega_1}\dots \hat a^\dagger_{\omega_N}|0\rangle\\
|N,N\rangle_\sigma
&=& {1\over {N!}}\left( {{\sigma}\over \pi}\right)^{N/2} \int d\omega_1\dots d\omega_{2N}
e^{-{\sigma\over 2} \sum_{j=1}^{2N}(\omega_j -\omega_0)^2}\nonumber\\
& &  \qquad \qquad  \times \hat a^\dagger_{\omega_1}\dots \hat a^\dagger_{\omega_N}
\hat b^\dagger_{\omega_{N+1}}\dots \hat b^\dagger_{\omega_{2N}}|0\rangle\end{eqnarray}
and \begin{eqnarray} & &
{1\over \sqrt{2}}\left[ |N,0\rangle +|0,N\rangle\right]_\sigma \nonumber \\
& & \quad ={1\over \sqrt{N!}}{1\over \sqrt{2}}\left( {{\sigma}\over \pi}
\right)^{N/4}\; \int d\omega_1\dots d\omega_N
e^{-{\sigma\over 2} \sum_{j=1}^N
(\omega_j -\omega_0)^2} \nonumber \\
& & \quad\qquad \times \left[ \hat a^\dagger_{\omega_1}\dots \hat a^\dagger_{\omega_N}
+\hat b^\dagger_{\omega_1}\dots \hat b^\dagger_{\omega_N}\right] |0\rangle ,\end{eqnarray}
where the bandwidth of the incident beams is given by $\Delta\omega\equiv \sigma^{-1/2}$.
If the photons are produced by SPDC, then the frequencies must occur in pairs with the photons in
each pair being equal distances above or below the pump frequency; we will consider this situation in
simplified form in section \ref{spdc} and in a more realistic form in section \ref{spdcreal}.

Suppose that one of the $N$-photon or $2N$-photon states described above is
input to the interferometer. Write this input state
as $|\psi_{in}\rangle $. Then, assuming that the frequencies of the
final photons are not measured, we want the joint probabilities to find $N_c$
photons at detector $C$ and $N_d$ photons at detector $D$ (with $N_c+N_d
=N_a+N_b$) for a given nondispersive phase shift $\phi_0$ in the
upper interferometer arm. These probabilities can
be expressed in the form \begin{equation} P(N_c,N_d|\phi_0)=\langle
\psi_{in}|\hat \pi (N_c,N_d;\phi_0)|\psi_{in}\rangle ,\end{equation} where the
projective operator
$\hat \pi(N_c,N_d;\phi_0)$ is defined as \begin{equation} \hat \pi
(N_c,N_d,\phi_0)=\int \; d\Omega |N_c,N_d;\Omega ,\phi_0\rangle \langle N_c,N_d;
\Omega ,\phi_0| ,\end{equation} with \begin{equation} |N_c,N_d;\Omega
,\phi_0\rangle ={1\over \sqrt{N_c!\; N_d!}} \hat c_{\Omega_1}^\dagger \dots
\hat c_{\Omega_{N_c}}^\dagger\hat d_{\Omega_1^\prime }^\dagger\dots
\hat d_{\Omega_{N_d}^\prime }^\dagger |0\rangle .\end{equation} Here we have
suppressed the $\phi_0$-dependence of the $\hat c_\Omega$ and $\hat
d_{\Omega^\prime}$ operators for notational simplicity, and have
represented the collection of output frequencies $\left\{ \Omega_1,\dots ,
\Omega_{N_c},\Omega_1^\prime ,\dots ,\Omega_{N_d}^\prime \right\} $ by the single
symbol $\Omega$. Similarly, $d\Omega$ is being used as shorthand for the full frequency
integration measure $d\Omega_1\dots d\Omega_{N_c}d\Omega_1^\prime \dots d\Omega_{N_d}^\prime $.
These probabilities may also be expressed in the form \begin{equation}
P(N_c,N_d|\phi_0)=\int d\Omega \left|\langle N_c,N_d;\Omega ,\phi_0|\psi_{in}\rangle \right|^2.\end{equation}
From equation \ref{info} the mutual information
between the phase $\Phi$ and the output photon numbers $M$ is then \begin{widetext}\begin{equation}H(\Phi :M) ={1\over {2\pi}} \sum_{N_c,N_d} \int_{-\pi}^\pi
d\phi_0 \; P(N_c,N_d|\phi_0) \log_2 \left[ {{2\pi P(N_c,N_d|\phi_0)}
\over {\int_{-\pi}^\pi d\phi_0\; P(N_c,N_d|\phi_0) }}\right] .\end{equation}\end{widetext}
We note that the probabilities $P(N_c,N_d,|\phi_0)$ are also conditional upon
the values of $\alpha$, $\beta$, and $\sigma$, although we do
not explicitly include these parameters in the notation for the probabilities
for the sake of notational simplicity.
We now restrict ourselves to the cases $N=1$ and $N=2$, and proceed in the following sections to compute the mutual
information $H$ for a number of different possible input states.

\section{MZ Interferometry With One-Photon Input}\label{mz1}

In this section, we begin with the cases in which there is only one photon in the initial state.

{\bf Case A: One-photon Fock state.} We introduce the normalized input state
\begin{equation}|\psi_{in} \rangle_\sigma =|10\rangle_\sigma
=\sqrt[4]{\sigma\over \pi}\int d\omega\;
e^{-{1\over 2}\sigma (\omega-\omega_0)^2}\hat a_\omega^\dagger |0\rangle ,\end{equation}
representing a single photon
incident on port A. Using relations \ref{adef} and \ref{bdef}, this is
equivalent to {\small
\begin{eqnarray}|10 \rangle_\sigma &=&{1\over 2}\sqrt[4]{\sigma\over \pi}\int d\omega\;
e^{-{1\over 2}\sigma (\omega-\omega_0)^2}\nonumber \\ & & \times \left[ \hat c^\dagger_\omega
\left( e^{i\phi (\omega )}-1\right) -i\hat d^\dagger_\omega
\left( e^{i\phi (\omega )}+1\right)\right] |0\rangle .\end{eqnarray}}

The photon may leave the interferometer
via either port C or port D.
We assume that the detectors count the number of
photons leaving the apparatus but do not measure their frequencies. Therefore, we must integrate over the
final frequencies.
The output state is then measured using the projective operators
\begin{eqnarray} \hat \pi (1,0) &=& \int d\Omega\; \hat c_\Omega^\dagger |0\rangle \langle 0|\hat c_\Omega\\
\hat \pi (0,1) &=& \int d\Omega \; \hat d_\Omega^\dagger |0\rangle \langle 0|\hat d_\Omega .\end{eqnarray}

\begin{figure}
\centering
\includegraphics[totalheight=2in]{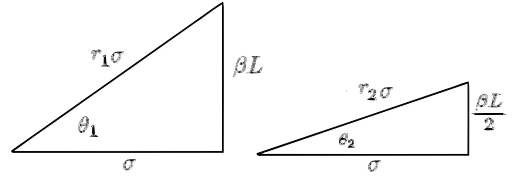}
\caption{Definitions of $r_1$, $r_2$, $\theta_1$, and $\theta_2$.}
\label{trianglefig}
\end{figure}

Expectation values of these operators give the probabilities of measurement outcomes:
{\footnotesize \begin{eqnarray}
P(1,0|\phi_0 )&=&{1\over 2}\left[ 1-{{e^{-{{\alpha^2L^2}\over {4r_1^2\sigma}}}}\over \sqrt{r_1}}\cos \left(
\phi_0 +{\theta_1\over 2}-{{\alpha^2\beta L^3}\over {4r_1^2\sigma^2}}\right)
\right] \label{probA1}\\
P(0,1|\phi_0 )&=&{1\over 2}\left[ 1+{{e^{-{{\alpha^2L^2} \over
{4r_1^2\sigma}}}}\over \sqrt{r_1}}\cos \left( \phi_0
+{\theta_1\over 2}-{{\alpha^2\beta L^3}\over {4r_1^2\sigma^2}}
\right) \right] \label{probA2} \end{eqnarray}} In the previous two
lines, we have introduced some notation that will be convenient
for simplifying the results of this and the following sections.
The parameters $r_1,r_2,\theta_1, \theta_2$ are defined by (see
figure \ref{trianglefig}):
\begin{eqnarray}r_1^2 &=& 1+\left({{\beta L}\over \sigma}\right)^2\qquad
\qquad \tan \theta_1\; =\; {{\beta L}\over \sigma}\\
r_2^2 &=& 1+\left({{\beta L}\over {2\sigma}}\right)^2
\qquad \qquad \tan \theta_2\; =\; {{\beta L}\over {2\sigma}}.
\end{eqnarray} Note that these
parameters depend on the second order dispersion coefficient $\beta$, but not on $\alpha$, and that when
$\beta $ vanishes we then have $r_1=r_2=1$
and $\theta_1=\theta_2=0$.

{\bf Case B: One-photon N00N state.} The input state is {\small \begin{equation}{1\over \sqrt{2}}\left[ |10\rangle
+|01\rangle \right]_\sigma =
{1\over \sqrt{2}}\sqrt[4]{\sigma\over \pi}\int d\omega\;
e^{-{\sigma\over 2}(\omega -\omega_0)^2}
\left( \hat a_\omega^\dagger +\hat b_\omega^\dagger \right) |0\rangle ,\end{equation}}
where
\begin{eqnarray} {1\over \sqrt{2}}\left( \hat a_\omega^\dagger
+\hat b_\omega^\dagger\right) &=& {i\over {2\sqrt{2}}} \left\{ \hat c^\dagger_\omega \left[ (i-1)-(i+1)e^{i\phi (\omega )}\right]
\right. \\ & & \left. \qquad +\; \hat d^\dagger_\omega \left[ e^{i\phi
(\omega )} (i-1) -(i+1)\right] \right\} .\nonumber \end{eqnarray}
The resulting output probabilities in this case turn out to be
\footnotesize{\begin{eqnarray}
P(1,0|\phi_0 )&=&{1\over 2}\left[ 1-{{e^{-{{\alpha^2\beta L^3}\over {4r_1^2\sigma}} }}\over \sqrt{r_1}}\sin \left(
\phi_0 +{\theta_1\over 2}-{{\alpha^2\beta L^3}\over {4r_1^2\sigma^2}}
\right)\right] \\
P(0,1|\phi_0 )&=&{1\over 2}\left[ 1+{{e^{-{{\alpha^2\beta L^3}\over {4r_1^2\sigma}}}}\over \sqrt{r_1}}\sin \left(
\phi_0 +{\theta_1\over 2}-{{\alpha^2\beta L^3}\over {4r_1^2\sigma^2}}\right)
\right] .\end{eqnarray}}

\section{MZ Interferometry With Two-Photon Input}\label{mz2}

We now consider input states with two photons distributed in assorted ways among the input ports. However, now we must
make a distinction as to whether the two photon frequencies are independent or correlated in some manner. We treat the uncorrelated
version first. Then we will examine one particular case of frequency-correlated photons which is of special interest for
experiment: that of photon pairs created through spontaneous parametric downconversion (SPDC). In this section
we treat only a simplified version of SPDC which will allow us to obtain simple exact expressions for the probabilities
of all of the output states. In a later section we will compare this simplified SPDC to a more realistic version for which
only numerical results are available.

\subsection{Two-Photon Input with Uncorrelated Energies}\label{mzun}

{\bf Case C: Energy-uncorrelated two-photon Fock state.}
Sending a two-particle Fock state into input A,
{\small \begin{equation}|2,0\rangle_\sigma = \sqrt{\sigma\over {2\pi}}
\int d\omega_1 d\omega_2 \; e^{-{\sigma\over 2}\left[ (\omega_1-\omega_0)^2
+(\omega_2-\omega_0)^2\right]}\hat a^\dagger_{\omega_1}\hat
a^\dagger_{\omega_2}|0\rangle,\end{equation}}
where we use equations \ref{adef} and \ref{bdef} to write each $\hat a^\dagger$ factor in terms of the
output operators $\hat c^\dagger$ and $\hat d^\dagger$.
After a straightforward calculation, this leads to the following
output probabilities:

{\footnotesize \begin{eqnarray}
P(2,0|\phi_0 )&=& {1\over 4} \left[ 1-{{e^{-{{\alpha^2L^2}\over {4r_1^2\sigma}}}}\over \sqrt{r_1}}
\cos \left( \phi_0+{{\theta_1}\over 2}-{{\alpha^2\beta L^3}\over
{4r_1^2\sigma^2}}\right)  \right]^2\\
P(0,2|\phi_0 )&=& {1\over 4} \left[ 1+{{e^{-{{\alpha^2L^2}\over {4r_1^2\sigma}}} }\over \sqrt{r_1}}
\cos \left( \phi_0+{{\theta_1}\over 2}-{{\alpha^2\beta L^3}\over
{4r_1^2\sigma^2}}\right) \right]^2\\
P(1,1|\phi_0 )&=& {1\over 2} \left[ 1-{{e^{-{{\alpha^2L^2}\over {2r_1^2\sigma}}}}\over {r_1}}\cos^2
\left( \phi_0+{{\theta_1}\over 2}
-{{\alpha^2\beta L^3}\over
{4r_1^2\sigma^2}}\right)  \right]
\end{eqnarray}}


{\bf Case D: Energy-uncorrelated two-photon dual Fock input.} The normalized input state is
\begin{widetext}\begin{eqnarray}|1,1\rangle_\sigma &=& \sqrt{\sigma\over \pi}
\int d\omega_1\omega_2\; e^{-{\sigma\over 2} \left[ (\omega_1-\omega_0)^2
+(\omega_2-\omega_0)^2\right] }\hat a_{\omega_1}^\dagger \hat a_{\omega_2}^\dagger |0\rangle \\
&=& {1\over 4}\sqrt{\sigma\over \pi}
\int d\omega_1\omega_2\; e^{-{\sigma\over 2} \left[ (\omega_1-\omega_0)^2
+(\omega_2-\omega_0)^2\right] }\\ & & \times \left\{ -i\hat
c_{\omega_1}^\dagger \hat c_{\omega_2}^\dagger
\left( e^{i\phi (\omega_1)}-1\right) \left( e^{i\phi (\omega_2)}+1\right)
+i\hat d_{\omega_1}^\dagger \hat d_{\omega_2}^\dagger
\left( e^{i\phi (\omega_2)}-1\right) \left( e^{i\phi (\omega_1)}+1\right)
\right. \nonumber \\
& & \qquad \left.  +\hat c_{\omega_1}^\dagger \hat d_{\omega_2}^\dagger
\left( e^{i\phi (\omega_1)}-1\right)\left( e^{i\phi (\omega_2)}-1\right)
-\hat c_{\omega_2}^\dagger \hat d_{\omega_1}^\dagger
\left( e^{i\phi (\omega_1)}+1\right)\left( e^{i\phi
(\omega_2)}+1\right) \right\} |0\rangle , \nonumber \end{eqnarray}\end{widetext}
which gives the results {\small \begin{eqnarray}P(2,0)&=& P(0,2) \nonumber \\
&=& {1\over 4} \left\{ 1-{{e^{-{{\alpha^2L^2}\over { 2r_1^2\sigma}}}}
\over {r_1}}
\cos \left[ 2\phi_0+\theta_1-{{\alpha^2\beta L^3}\over
{2r_1^2\sigma^2}}\right]\right\} \label{iv1}\\
P(1,1) &=& {1\over 2} \left\{ 1+{{e^{-{{\alpha^2L^2}\over { 2r_1^2\sigma}}}}
\over {r_1}}
\cos \left[ 2\phi_0+\theta_1-{{\alpha^2\beta L^3}\over
{2r_1^2\sigma^2}}\right]\right\} \label{iv2}\end{eqnarray}}

{\bf Case E: Energy-uncorrelated two-photon N00N state.}

For the input state \begin{eqnarray} |20\rangle_\sigma +|02\rangle_\sigma
&=& \sqrt{\sigma\over {2\pi}}\int d\omega_1\; d\omega_2
e^{-{\sigma\over 2}[(\omega_1-\omega_0)^2+(\omega_2-\omega_0)^2]}\nonumber\\
& & \qquad \times {1\over \sqrt{2}}\left( \hat
a_{\omega_1}^\dagger \hat a_{\omega_2}^\dagger +\hat
b_{\omega_1}^\dagger \hat b_{\omega_2}^\dagger \right) |0\rangle ,
\end{eqnarray} the output probabilities are
\begin{eqnarray}P(2,0|\phi_0 )&=&  P(0,2|\phi_0 ) \; =\; {1\over 4}\left\{
1+{1\over {r_1}} e^{-{{\alpha^2L^2}\over {2r_1^2 \sigma}}}
\right\} \\
P(1,1|\phi_0 ) &=& {1\over 2}\left\{ 1-{1\over {r_1}} e^{-{{\alpha^2L^2}
\over {2r_1^2 \sigma}}}\right\} .\end{eqnarray} In the absence of
dispersion ($\alpha=\beta =0$) or in the narrow bandwidth limit ($\sigma \to \infty$),
we see that the coincidence rate $P(1,1|\phi_0 )$ vanishes,
while the other two probabilities are both equal to ${1\over 2}$.

Note that there is no dependence on $\phi_0$. We will see later that this
fact manifests itself in a vanishing mutual information.

\subsection{Two-Photon Input with Anticorrelated Energies:
Simplified SPDC model}\label{spdc}

{\bf Case F: Simplified SPDC Fock states.}
We now examine a case with two photons incident on the same
input port and anticorrelated in energy. We do this
in the context of a simplified model of spontaneous
parametric downconversion (SPDC). Energy conservation requires that the two downconverted photons have
frequencies $\omega_\pm =\omega_0\pm \Omega $, where $2\omega_0 $ is the pump frequency.
We again assume a Gaussian distribution of frequencies, centered around
$\omega_0$, of the
form $e^{-{1\over 2}\sigma (\omega_\pm -\omega_0)^2}=e^{-{\sigma\over 2}
\Omega^2}$. We follow essentially the same
calculational procedure as before, except now we enforce the requirement
that the incoming photon frequencies satisfy $\omega_1+\omega_2=2\omega_0$.
In this section we impose this condition in a manner that will allow us
to obtain analytic solutions for the output probabilities. This will serve
us as a simplified version of SPDC, and we will see in section~\ref{spdcreal}
that this model seems to give an upper bound to the mutual information obtained from
a more realistic model of SPDC. The input state in this model is taken to be of the form
\begin{equation}|20\rangle_\sigma =\sqrt[4]{{2\sigma}\over \pi}
\int_{-\infty}^{\infty}d\Omega \int_{-\infty}^\infty d\epsilon\;
e^{-\sigma \Omega^2} f (\epsilon ) a_{\omega_+}^\dagger
a_{\omega_-}^\dagger |0\rangle ,\end{equation} where now
$\omega_+=\omega_0+\Omega $ and
$\omega_-=\omega_0-\Omega+\epsilon$. We can choose $f(\epsilon )$
to be any function sharply peaked at zero with normalized integral
(unit area under its graph). We then compute the output
probabilities according to
\begin{equation}P(N_c,N_d|\phi_0 )= \int d\omega_1 d\omega_2
\left| \langle N_c,N_d|\psi_{in}\rangle \right|^2 ,\end{equation}
or equivalently, by applying the projection operators
\begin{equation} \hat \pi (N_c,N_d) =\int d\omega_1\; d\omega_2
|N_c,N_d\rangle \langle N_c,N_d| .\end{equation} The auxiliary
function $f_\lambda (\epsilon )$ is necessary in this model in
order to impose the constraint $\omega_1+\omega_2=2\omega_0$
without causing squares of delta functions to arise in the
probability calculations. A more correct treatment of SPDC will
follow in section~\ref{spdcreal}.

The measurement outcomes, integrated over final frequency, are then given
by \begin{eqnarray}
P(2,0|\phi_0 )&=& {1\over 8}\left\{ 2+e^{-{{\alpha^2L^2}\over
{2\sigma}}}+{1\over \sqrt{r_1}} \cos \left( 2\phi_0 +{{\theta_1}\over
2}\right) \right. \\
& & \qquad \left. -{4\over \sqrt{r_2}} e^{-{{\alpha^2L^2}\over {8r_2^2\sigma}}}
\cos \left[ \phi_0 +{{\theta_2}\over 2}-
{{\alpha^2\beta L^3}\over {16r_2^2\sigma^2}}\right] \right\}\nonumber \\
P(0,2|\phi_0 ) &=& {1\over 8}\left\{ 2+e^{-{{\alpha^2L^2}\over
{2\sigma}}}+{1\over \sqrt{r_1}} \cos \left( 2\phi_0 +{{\theta_1}\over
2}\right) \right. \\
& & \qquad \left. +{4\over \sqrt{r_2}} e^{-{{\alpha^2L^2}\over {8r_2^2\sigma}}}
\cos \left[ \phi_0 +{{\theta_2}\over 2}-
{{\alpha^2\beta L^3}\over {16r_2^2\sigma^2}}\right]\right\} \nonumber \\
P(1,1|\phi_0 ) &=& {1\over 4}\left\{ 2-e^{-{{\alpha^2L^2}\over
{2\sigma}}}-{1\over \sqrt{r_1}} \cos \left( 2\phi_0 +{{\theta_1}\over
2}\right)\right\} .\end{eqnarray}
As $\beta $ increases, $r_1$ and $r_2$ increase, leading to decreased visibility of all of the
oscillating terms.

Note also that in the case of zero dispersion ($\alpha =\beta =0$), the exact expressions for energy-uncorrelated (Case C, section 4.1)
and energy-anticorrelated (downconverted) Fock states (Case F) are identical to each other. However the probabilities begin to diverge when
dispersion is turned on. The same effect will be seen to occur for the uncorrelated and anticorrelated N00N states in the HOM interferometer
(cases J and K, below).

{\bf Case G: Simplified SPDC N00N states.} Now the input state is taken to be a N00N state,
${1\over \sqrt{2}} \left\{ |20\rangle_{\lambda ,\sigma} +|02\rangle_{\lambda ,\sigma}
\right\} $. We find the measurement outcomes to be:
\begin{eqnarray}
P(2,0|\phi_0 )&=& P(0,2) \; =\; {1\over 4}
\left[ 1+e^{-{{\alpha^2 L^2}\over {2\sigma}}}\right] \\
P(1,1|\phi_0 )&=& {1\over 2} \left[ 1-e^{-{{\alpha^2L^2}\over
{2\sigma}}}\right]
\end{eqnarray}

As in the uncorrelated case, the probabilities show no dependence on $\phi_0$,
and so have vanishing mutual information.In this case we also see that there is no dependence on
the 2nd order dispersion coefficient $\beta$.

It is interesting to note what happens if we shift the phase of
the photons in one input port by $\pi\over 2$ before they
hit the first beamsplitter. The input to the interferometer is now
proportional to
$\left| 2,0\rangle \right. -\left| 0,2 \rangle \right.  .$ In this
case, the interference in $\phi_0$ reemerges, and the result is
independent of $\alpha$ instead of $\beta$. In fact, the
counting probabilities turn out to be very similar to those of the
N00N state incident on an HOM interferometer presented in the next section (case K).
Moreover, these two cases have identical values for the mutual information.

{\bf Case H: Simplified SPDC dual Fock state.} The frequency-anticorrelated dual Fock input state
\begin{equation}|1,1\rangle_\sigma = \sqrt[4]{\sigma\over {2\pi}}
\int d\Omega\int d\epsilon\; e^{-\sigma\Omega^2}f(\epsilon )\hat a_{\omega_+}^\dagger \hat b_{\omega_-}^\dagger |0\rangle
\end{equation}
gives the results \begin{eqnarray}P(2,0|\phi_0 )&=& P(0,2|\phi_0 ) \\ \nonumber
&=& {1\over 4} \left[ 1-{1\over \sqrt{r_1}}\cos \left(
2\phi_0 +{{\theta_1}\over 2}\right) \right] \label{H1}\\
P(1,1|\phi_0 ) &=& {1\over 2} \left[ 1+{1\over \sqrt{r_1}}\cos \left(
2\phi_0 +{{\theta_1}\over 2}\right) \right] . \label{H2}\end{eqnarray}

\section{Dispersive Hong-Ou-Mandel Interferometer With N00N Input}\label{hom}

An alternative setup has been proposed to improve phase
resolution(\cite{dowl}). In this section we examine this alternate
version and compare it to the previous results.

\begin{figure}
\centering
\includegraphics[totalheight=2in]{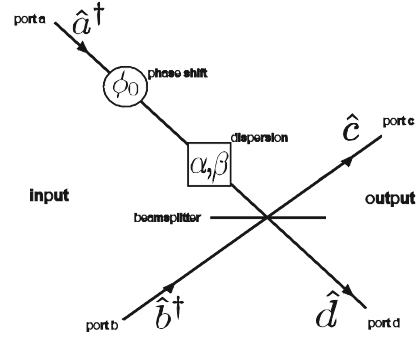}
\caption{Hong-Ou-Mandel interferometer with dispersion and non-dispersive
phase shift $\phi_0$ in one arm.}
\label{homfig}
\end{figure}

In this version, it is assumed that the N00N state is created {\it inside} the interferometer, rather than
at the input ports. Effectively, we need to remove the first beam splitter
from the interferometer and use the N00N state as input to the remaining
beamsplitter, which now acts as a Hong-Ou-Mandel (HOM) interferometer~\cite{hompaper}.
The setup is shown in figure \ref{homfig}. We again assume dispersion and phase-shift $\phi_0$ along
one of the lines entering the beam splitter, neglecting absorption.
Ignoring an overall constant phase of $e^{ik_0L}$, the scattering matrix
now has the form \begin{equation}S(\phi_0) = {1\over \sqrt{2}}
\left( \begin{array}{cc} e^{i\phi (\omega)} & i\\
ie^{i\phi (\omega)} & +1\end{array}\right) ,\end{equation} where $\phi (\omega )$
is again given by equation \ref{phiomega}.
Thus, \begin{eqnarray}\hat c_\omega&=& {1\over \sqrt{2}}\left[
e^{i\phi (\omega )}
\hat a_\omega +i \hat b_\omega\right] \\
\hat d_\omega &=& {1\over \sqrt{2}}\left[ i e^{i\phi (\omega )}
\hat a_\omega + \hat b_\omega\right] .\end{eqnarray}

{\bf Case I: Single-photon N00N state in HOM interferometer.}
By the same methods as before, we can compute the counting rates for a given N00N state input.
The one-photon N00N input state is
\begin{equation} |\psi \rangle =C\int d\omega \; e^{-{\sigma\over 2}(\omega -\omega_0)^2 }
(a_\omega^\dagger +b_\omega^\dagger ) |0\rangle ,\end{equation}
for which we find {\footnotesize \begin{eqnarray}
P(1,0|\phi_0 )&=& {1\over 2} \left\{ 1+{{e^{-{{\alpha^2L^2}\over { 4r_1^2\sigma}}}}\over \sqrt{r_1}}
\sin \left[ \phi_0+{\theta_1\over 2}
-{{\alpha^2\beta L^3}\over
{4r_1^2\sigma^2}} \right]\right\} \\
P(0,1|\phi_0 )&=& {1\over 2} \left\{ 1-{{e^{-{{\alpha^2L^2}\over { 4r_1^2\sigma}}}}\over \sqrt{r_1}}
\sin \left[ \phi_0+{\theta_1\over 2}
-{{\alpha^2\beta L^3}\over
{4r_1^2\sigma^2}} \right]\right\} \end{eqnarray}}

Finally, assuming two-photons, with no correlation or with anticorrelation,
we arrive at two additional cases (J and K).

{\bf Case J: Energy-uncorrelated two-photon N00N state in HOM interferometer.}

The input state is {\small
\begin{equation} |\psi \rangle =C\int d\omega_1 d\omega_2 \; e^{-{\sigma\over 2}\left[
(\omega_1 -\omega_0)^2 +(\omega_2-\omega_0)^2\right] }
(a_{\omega_1}^\dagger a_{\omega_2}^\dagger +b_{\omega_1}^\dagger b_{\omega_2}^\dagger  ) |0\rangle .\end{equation}}
From this state, we arrive at the results:
{\footnotesize \begin{eqnarray}
P(2,0|\phi_0 )&=& P(0,2|\phi_0 ) \nonumber \\ &=& {1\over 4}
\left[ 1- {{e^{-{{\alpha^2L^2}\over { 2r_1^2\sigma}}}}\over {r_1}}
\cos \left( 2\phi_0+\theta_1
-{{\alpha^2\beta L^3}\over {2r_1^2\sigma^2}} \right) \right] \label{x1}\\
P(1,1|\phi_0 )&=& {1\over 2} \left[ 1+{{e^{-{{\alpha^2L^2}\over { 2r_1^2\sigma}}}}\over {r_1}}
\cos \left( 2\phi_0+\theta_1
-{{\alpha^2\beta L^3}\over {2r_1^2\sigma^2}} \right) \right] \label{x2}
\end{eqnarray}}

{\bf Case K: Simplified SPDC two-photon N00N state in HOM interferometer.}

For the input
\begin{equation} |\psi \rangle =C\int d\Omega \; e^{-\sigma
\Omega^2}
(a_{\omega_+}^\dagger a_{\omega_-}^\dagger +b_{\omega_+}^\dagger b_{\omega_-}^\dagger  ) |0\rangle ,\end{equation}
we compute:
\begin{eqnarray}
P(2,0|\phi_0 )&=& P(0,2|\phi_0 ) \nonumber \\ &=& {1\over 4}
\left[ 1- {1\over \sqrt{r_1}} \cos \left( 2\phi_0+{{\theta_1}\over 2}
\right) \right] \label{K1}\\
P(1,1|\phi_0 )&=& {1\over 2} \left[ 1+{1\over \sqrt{r_1}}
\cos\left( 2\phi_0+{{\theta_1}\over 2}\right) \right] \label{K2}
\end{eqnarray}

In this last case, the results turn out to be independent of the first
order dispersion coefficient, $\alpha$.

\section{Comparison and Discussion of Cases A to K} \label{comp}

\begin{figure}
\centering
\includegraphics[totalheight=2in]{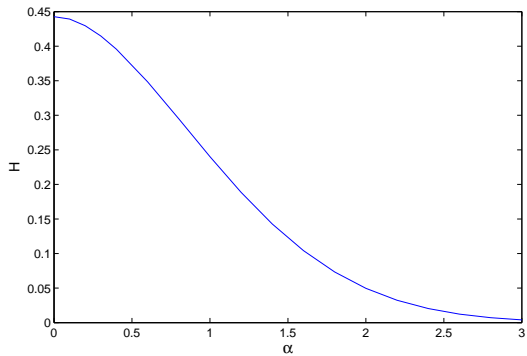}
\caption{(color online). Mutual information versus alpha for
single photon cases (cases A, B, and I), plotted for the values
$\beta =0$, $\sigma =1$. The mutual information is the same for
all three cases. ($\alpha$, $\beta$, and $\sigma$ are in units of
$L^{-1}\omega_0^{-1}$, $L^{-1} \omega_0^{-2}$, and
$\omega_0^{-2}$, respectively.)} \label{alphafig1}
\end{figure}

\begin{figure}
\centering
\includegraphics[totalheight=2in]{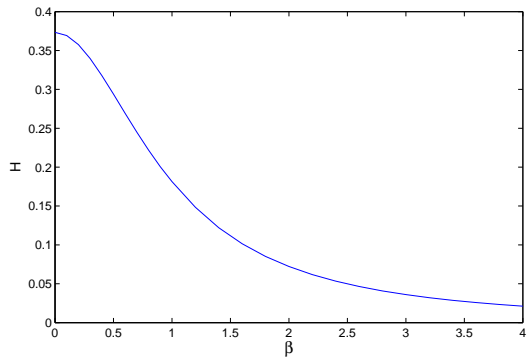}
\caption{(color online). Mutual information versus $\beta$ for
single photon cases (cases A, B, and I), for the values $\alpha
=.5$, $\sigma =1$. ($\alpha$, $\beta$, and $\sigma$ are in units
of $L^{-1}\omega_0^{-1}$, $L^{-1} \omega_0^{-2}$, and
$\omega_0^{-2}$, respectively.)} \label{betafig1}
\end{figure}

\begin{figure}
\centering
\includegraphics[totalheight=2in]{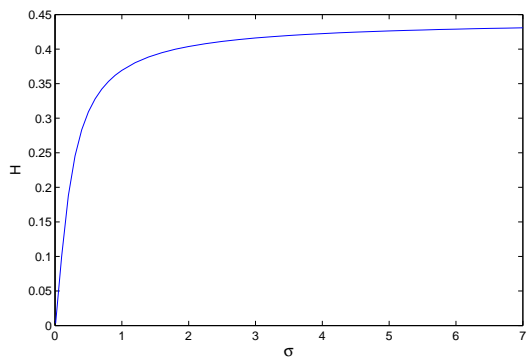}
\caption{(color online). Mutual information versus squared inverse
bandwidth $\sigma$ for single photon cases (cases A, B, and I),
for the values $\alpha =1$, $\beta =.1$. ($\alpha$, $\beta$, and
$\sigma$ are in units of $L^{-1}\omega_0^{-1}$, $L^{-1}
\omega_0^{-2}$, and $\omega_0^{-2}$, respectively.)}
\label{sigmafig1}
\end{figure}

\begin{figure}
\centering
\includegraphics[totalheight=2in]{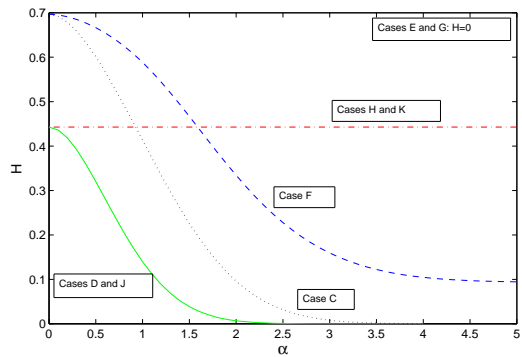}
\caption{(color online). Mutual information versus alpha for
two-photon cases, for the values $\sigma =1$, $\beta =0$. Cases E
and G vanish identically. Cases J and D are identical, as are
cases H and K. ($\alpha$, $\beta$, and $\sigma$ are in units of
$L^{-1}\omega_0^{-1}$, $L^{-1} \omega_0^{-2}$, and
$\omega_0^{-2}$, respectively.)} \label{alphafig2}
\end{figure}

\begin{figure}
\centering
\includegraphics[totalheight=2in]{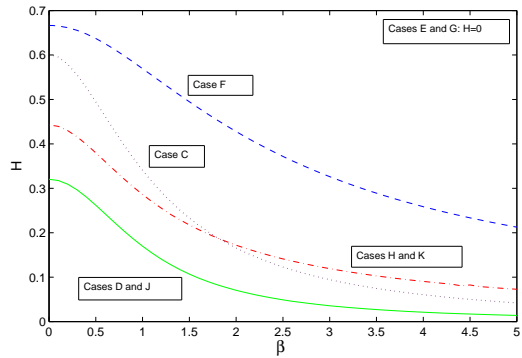}
\caption{(color online). Mutual information versus beta for
two-photon cases, for the values $\sigma=1$, $\alpha=.5$.
($\alpha$, $\beta$, and $\sigma$ are in units of
$L^{-1}\omega_0^{-1}$, $L^{-1} \omega_0^{-2}$, and
$\omega_0^{-2}$, respectively.)} \label{betafig2}
\end{figure}

\begin{figure}
\centering
\includegraphics[totalheight=2in]{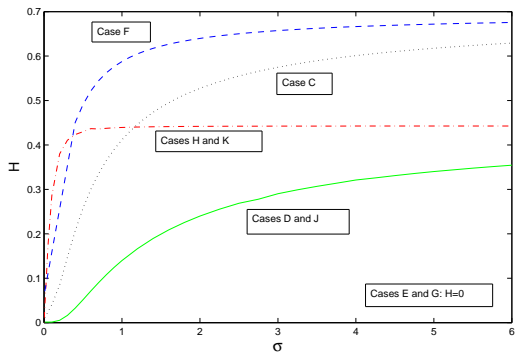}
\caption{(color online). Mutual information versus squared inverse
bandwidth $\sigma$ for two-photon cases, for the values $\alpha
=1$, $\beta =.1$. ($\alpha$, $\beta$, and $\sigma$ are in units of
$L^{-1}\omega_0^{-1}$, $L^{-1} \omega_0^{-2}$, and
$\omega_0^{-2}$, respectively.)} \label{sigmafig2}
\end{figure}

The detection probabilities of the previous sections can now be
combined with equation \ref{info} to compute the
mutual information for each of the experimental setups and inputs
states. Plotting the results as functions
of $\alpha$, $\beta$, and $\sigma$, we find the results in figures \ref{alphafig1}-\ref{sigmafig1} for
single photon
input and figures \ref{alphafig2}-\ref{sigmafig2} for two photons. $\alpha L$ is given in units of $\omega_0^{-1}$,
while $\beta L$ and $\sigma$ are in units of $\omega_0^{-2}$. In the dispersionless limit,
$\alpha ,\beta \to 0$, we find Shannon mutual information values that
agree with those previously calculated in \cite{bahdlap}.

Only positive values of $\beta$ were graphed. However, the formulas of the previous
sections work equally in the anomalous dispersion (negative $\beta$) region.

Note also that the four parameters $\alpha ,\beta ,\sigma ,L$ appear in all equations only through
the dimensionless quantities \begin{equation}
\Lambda_1 ={\sigma\over {\beta L}}\qquad
\qquad \mbox{and} \qquad \qquad \Lambda_2={\sigma\over {\alpha^2L^2}}
.\label{dimensionless}\end{equation}
Thus, other parameter ranges can easily be obtained from those graphed here via
appropriate rescaling of variables with the dimensionless ratios held fixed.

A few conclusions are immediately clear from these graphs and from
the equations of the previous sections. (i) First, the dual Fock
states entering the Mach-Zehnder interferometer give identical
results as the N00N states entering the Hong-Ou Mandel
interferometer (compare equations \ref{H1} and \ref{H2} to
\ref{K1} and \ref{K2}, or compare \ref{iv1} and \ref{iv2} to
\ref{x1} and \ref{x2}). This is to be expected, since for $N=2$ the two
cases are equivalent: the first beam splitter in the Mach-Zehnder
interferometer turns a dual Fock input state into a N00N output
state, which  then strikes the second beamsplitter. The second
beamsplitter can then be viewed as an HOM interferometer. Thus
cases J and D are equivalent, as are cases H and K. (This equivalence will
not for $N>2$.) (ii) Second,
the single-photon cases (cases A, B, and I) all give identical
curves for the mutual information as functions of $\alpha$,
$\beta$, and $\sigma$. The explanation for this is clear if the
action of the first beam splitter on the input is examined. Cases
B and I are equivalent for the same reason mentioned in the
previous point: they both lead to a one-particle N00N state in the
portion of the interferometer before the dispersive element is
reached, and so give the same output. Meanwhile, in case A, the
output of the first beam splitter is the state ${1\over
\sqrt{2}}\left( |01\rangle +i|10\rangle \right) ;$ this is similar
to a N00N state, except one term is shifted in phase by $\pi\over
2$ relative to the other. This converts the sines in the
probabilities of cases B and I into the cosines of case A
(equations \ref{probA1} and \ref{probA2}), but has no other
effect. Since the mutual information involves integrals from
$-\pi$ to $\pi$, interchanging sines and cosines inside the
integrals has no effect on the mutual information. Unsurprisingly,
the single photon cases generally result in lower mutual
information than the two-photon cases. (iii) We see from the
graphs for the two-photon states that the energy-uncorrelated and
energy-anticorrelated version of each input give identical results
for zero dispersion or zero bandwidth ($\sigma =\infty$);
however, the uncorrelated versions all drop off rapidly to zero fidelity as
the dispersion increases, whereas the anticorrelated
(downconverted) input leads to a much
slower drop.  (iv) Two-photon N00N states incident on
the MZ interferometer (cases E and G) have zero mutual information
as anticipated earlier. (v) For fixed bandwidth and fixed
quadratic dispersion coefficient $\beta$, the two-photon
downconverted N00N state in the HOM interferometer (case K) is
independent of the linear coefficient $\alpha$. However, it decays
rapidly with increasing $\beta$. (vi) Overall, the simplified
SPDC-generated Fock states (case F) seem to hold up best in the
presence of dispersion. This case starts off with a higher value
of $H$ at zero dispersion and decays more slowly as $\alpha$ and
$\beta$ increase. The only exception to this statement is when
$\beta$ is small, in which case the anticorrelated HOM N00N state
(case K) works better at large $\alpha$.

A bit of insight into some of the properties of the 2-photon results may be obtained by considering
the exponential decay factor \begin{equation}\zeta \equiv e^{-{{\alpha^2L^2}\over {4r_1^2\sigma}}}
=e^{-{{\Lambda_2}\over {4r_1^2}}}=e^{-{{\Lambda_2}\over {4(1+\Lambda_1^{-2})}}}.
\end{equation} $\Lambda_1$ and $\Lambda_2$ are the dimensionless
quantities defined in equation \ref{dimensionless}. In frequency-uncorrelated cases such as cases C and D, all of the $\phi_0$-dependent
terms are multiplied by a factor of $\zeta$ which arises from interference between
$e^{i\phi(\omega_1)}$ and $e^{i\phi(\omega_2)}$ terms, where $\omega_1$ and
$\omega_2$ are the frequencies of the photons entering the input ports. The
relevant
term is of the form $e^{i[\phi(\omega_1) +\phi (\omega_2)]}$. As $\alpha\to\infty$ or
$\sigma\to 0$, we find that $\Lambda_2\to 0$ and $\zeta\to 0$, so that only constant
($\phi_0$-independent) terms survive in the limit. Thus, for large $\alpha$ or small
$\sigma$, the dependence of the probability distributions on $\phi_0$ decays exponentially,
causing the mutual information to also decay rapidly.

In contrast, for the frequency-anticorrelated cases, such as F and H,
the term $e^{i(\phi(\omega_1) +\phi (\omega_2))}$ becomes \begin{equation}
e^{i[\phi(\omega_+)+\phi(\omega_-)]} =e^{i[2\phi_0 +2\beta \omega^2]},
\end{equation} with the $\alpha$-dependence
cancelling. As a result, $\phi_0$-dependent terms occur without the
exponentially decaying $\zeta$ factor, allowing much slower decay of $H$
at large $\alpha$ (or even no decay at all, as in case
H). The slower decay at large dispersion is therefore a direct
consequence of the quantum-mechanical correlations present in the initial
state.

As for the $\beta$ dependence, we see that as $\beta$ becomes large,
both $\zeta$ and $r_1$ become $\beta$ independent, with
$\zeta\to e^{-{{\Lambda_2}
\over 4}}$ and $r_1\to 1$;
thus all the curves approach constant
values at large $\beta$, with slopes ${{dH}\over {d\beta}}$
of comparable order of magnitude.

We turn now to one additional case, that of more realistic SPDC
photon pairs, which we then proceed to compare with the simplified
SPDC model already examined.

\section{Case L: SPDC}\label{spdcreal}

Now we present results for the mutual information using a more realistic
model for the parametric downconversion process. Numerically, the results
turn out qualitatively (and for some parameter ranges quantitatively as well)
to be very similar to those of the simplified SPDC model in the previous
section; however we no longer will be able to present explicit
analytic expressions for the measurement outcomes.

There are many possible cases that could be considered, but we restrict
ourselves here to the single case of collinear type-II SPDC in a nonlinear crystal,
with both of the outgoing photons entering port A of the dispersive
Mach-Zehnder interferometer. We now have to consider the parameters of both
the interferometer and the crystal. We allow the pump frequency to vary around
central frequency $2\omega_0$, with the deviation from the center of the
distribution represented by $2\Omega_p$; in other words, the
pump frequency is represented as $$\omega_p=2(\omega_0+\Omega_p).$$
We once again assume a Gaussian distribution of frequencies, in this
case represented by
a weighting factor $e^{-{{\sigma}\over 2}
(\omega_p-2\omega_0)^2}=e^{-2\sigma
\Omega_p^2}$. The signal and idler frequencies are then
\begin{eqnarray}
\omega_s &=& {{\omega_p}\over 2} +\Omega =\omega_0+\Omega_p+\Omega \\
\omega_i &=& {{\omega_p}\over 2} -\Omega =\omega_0+\Omega_p-\Omega ,
\end{eqnarray}
with $\omega_s+\omega_i=\omega_p$.  Suppose that the crystal is
cut so that exact phase matching occurs at the central frequency
\begin{equation} k_p(2\omega_0)=k_s(\omega_0) +k_i(\omega_0).
\end{equation} Then, assuming that terms quadratic
and higher in the frequencies are small, the phase matching condition for the
crystal gives us a condition on the wave-vectors of the form~\cite{ou}
\begin{equation}
\Delta k\equiv k_p(\omega_p) -k_s(\omega_s)-k_i(\omega_i) =\Lambda_p\Omega_p
+\Lambda\Omega ,\end{equation} where $\Lambda_p = 2k_p^\prime (2\omega_0)
-k_s^\prime (\omega_0) -k_i^\prime (\omega_0) $
and $\Lambda = k_i^\prime (\omega_0) -k_s^\prime (\omega_0) $.

The wavefunction for the biphoton state entering the interferometer is now
\begin{equation} |\psi_{in}\rangle=\int d\Omega \; d\Omega_p
\Phi(\Omega_p,\Omega ) \hat a_{\omega_0+\Omega_p+\Omega}^\dagger
\hat a_{\omega_0+\Omega_p-\Omega}^\dagger |0\rangle ,\label{spdcwavefunction}\end{equation}
where \begin{equation} \Phi(\Omega_p,\Omega ) = {\cal N}
e^{-2\sigma \Omega_p^2}\left( {{\sin{{\Delta kL_c}\over 2}}\over
{{\Delta kL_c}\over 2}} \right) e^{-i\Delta k\; L_c/2} ,\label{spdcprefactor}\end{equation}
with normalization constant ${\cal N}$. Here, $L_c$ is the length of the
nonlinear crystal. Using this wavefunction, we can compute
output probabilities as before.
Denoting the frequencies at the detectors by \begin{widetext}
$\omega$ and $\omega^\prime$, we have \begin{eqnarray}
P(2,0|\phi_0)&=& \int d\omega\; d\omega^\prime \left|
\Phi ({{\omega-\omega^\prime}\over 2}, \omega_0-{{\omega +\omega^\prime}
\over 2})+\Phi (-{{\omega-\omega^\prime}\over 2}, \omega_0-{{\omega +\omega^\prime}
\over 2}) \right|^2 \nonumber \\
& & \times \left| 1+e^{i[\phi (\omega )+\phi (\omega^\prime )]}
-e^{i\phi ( \omega )}-e^{i\phi (\omega^\prime )}\right|^2 \\
P(1,1|\phi_0)&=& \int d\omega\; d\omega^\prime \left|
\Phi ({{\omega-\omega^\prime}\over 2}, \omega_0-{{\omega +\omega^\prime}
\over 2})\left[ 1-e^{i[\phi (\omega )+\phi (\omega^\prime )]}
-e^{i\phi ( \omega )}+e^{i\phi (\omega^\prime )} \right. \right. \nonumber\\
& &+ \left.
\Phi ({{\omega-\omega^\prime}\over 2}, \omega_0-{{\omega +\omega^\prime}
\over 2})\left[ 1-e^{i[\phi (\omega )+\phi (\omega^\prime )]}
+e^{i\phi ( \omega )}-e^{i\phi (\omega^\prime )}\right] \right|^2 \\
P(0,2|\phi_0)&=& \int d\omega\; d\omega^\prime \left|
\Phi ({{\omega-\omega^\prime}\over 2}, \omega_0-{{\omega +\omega^\prime}
\over 2})+\Phi (-{{\omega-\omega^\prime}\over 2}, \omega_0-
{{\omega +\omega^\prime} \over 2}) \right|^2 \nonumber \\
& & \times  \left| 1+e^{i[\phi (\omega )+\phi (\omega^\prime )]}
+e^{i\phi ( \omega )}+e^{i\phi (\omega^\prime )}\right|^2 ,
\end{eqnarray}\end{widetext}
where $\phi (\omega )$ is as defined in equation~\ref{phiomega}.
Note that $\Phi \left({{\omega-\omega^\prime}\over 2},
\omega_0-{{\omega +\omega^\prime}\over 2}\right)$ depends only on
the crystal properties, while $\phi (\omega )$ depends only on the
properties of the interferometer. The integrands of
$P(0,2|\phi_0)$ and $P(2,0|\phi_0)$ factor in their dependence on
these two sets of parameters; that of $P(1,1|\phi_0)$ does not,
indicating the entangled nature of the $|11\rangle $ state.

\begin{figure}
\centering
\includegraphics[totalheight=2in]{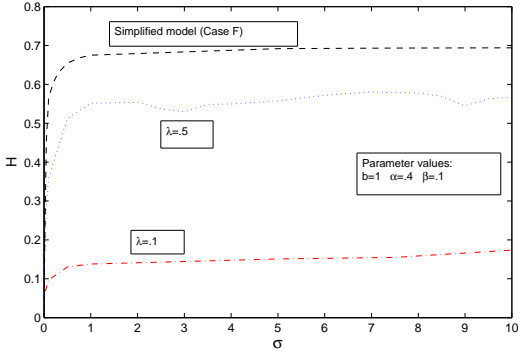}
\caption{(color online). Mutual information versus squared inverse
bandwidth $\sigma$ for SPDC. ($\alpha$, $\beta$, and $\sigma$ are
in units of $L^{-1}\omega_0^{-1}$, $L^{-1} \omega_0^{-2}$, and
$\omega_0^{-2}$, respectively. $b$ is in units of $\omega_0^{-2}$,
while $\lambda$ is dimensionless.)} \label{sigmafigx}
\end{figure}

\begin{figure}
\centering
\includegraphics[totalheight=2in]{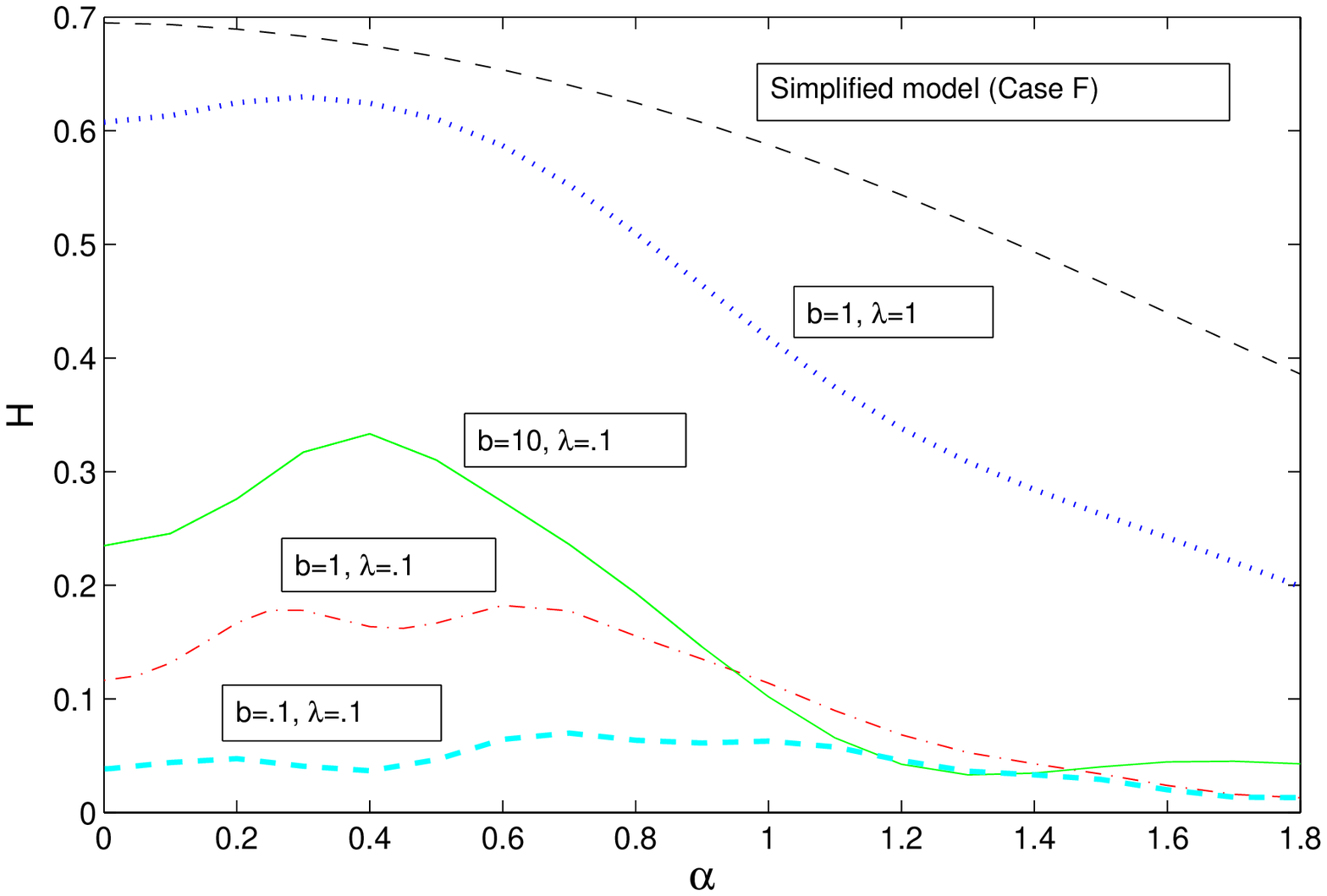}
\caption{(color online). Mutual information versus alpha for SPDC
with $\sigma=1$, $\beta=.1$. ($\alpha$, $\beta$, and $\sigma$ are
in units of $L^{-1}\omega_0^{-1}$, $L^{-1} \omega_0^{-2}$, and
$\omega_0^{-2}$, respectively. $b$ is in units of $\omega_0^{-2}$,
while $\lambda$ is dimensionless.)} \label{alphafigx}
\end{figure}

\begin{figure}
\centering
\includegraphics[totalheight=2in]{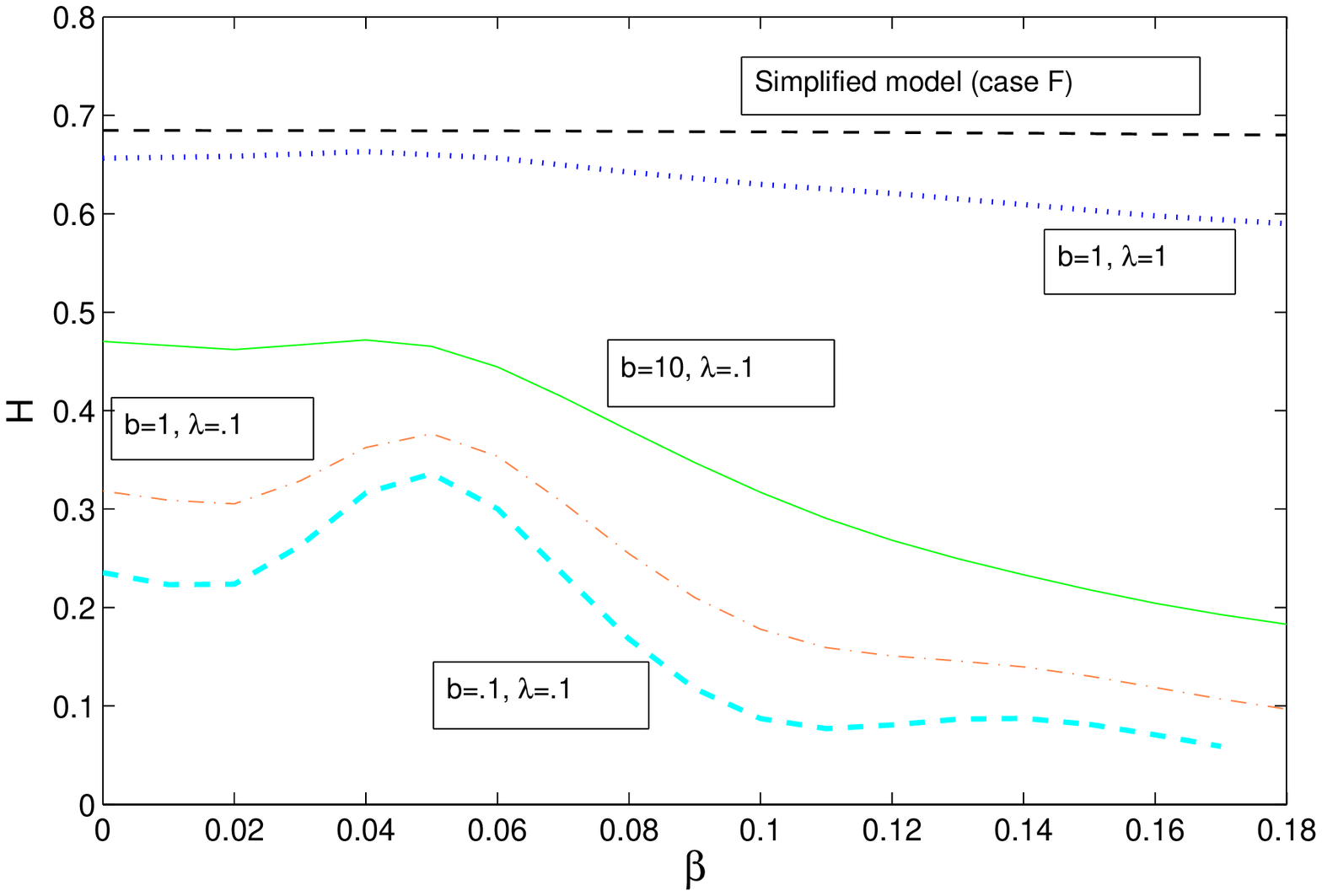}
\caption{(color online). Mutual information versus beta for SPDC
with $\sigma=1$, $\alpha=.3$. ($\alpha$, $\beta$, and $\sigma$ are
in units of $L^{-1}\omega_0^{-1}$, $L^{-1} \omega_0^{-2}$, and
$\omega_0^{-2}$, respectively. $b$ is in units of $\omega_0^{-2}$,
while $\lambda$ is dimensionless.)} \label{betafigx}
\end{figure}

\begin{figure}
\centering
\includegraphics[totalheight=2in]{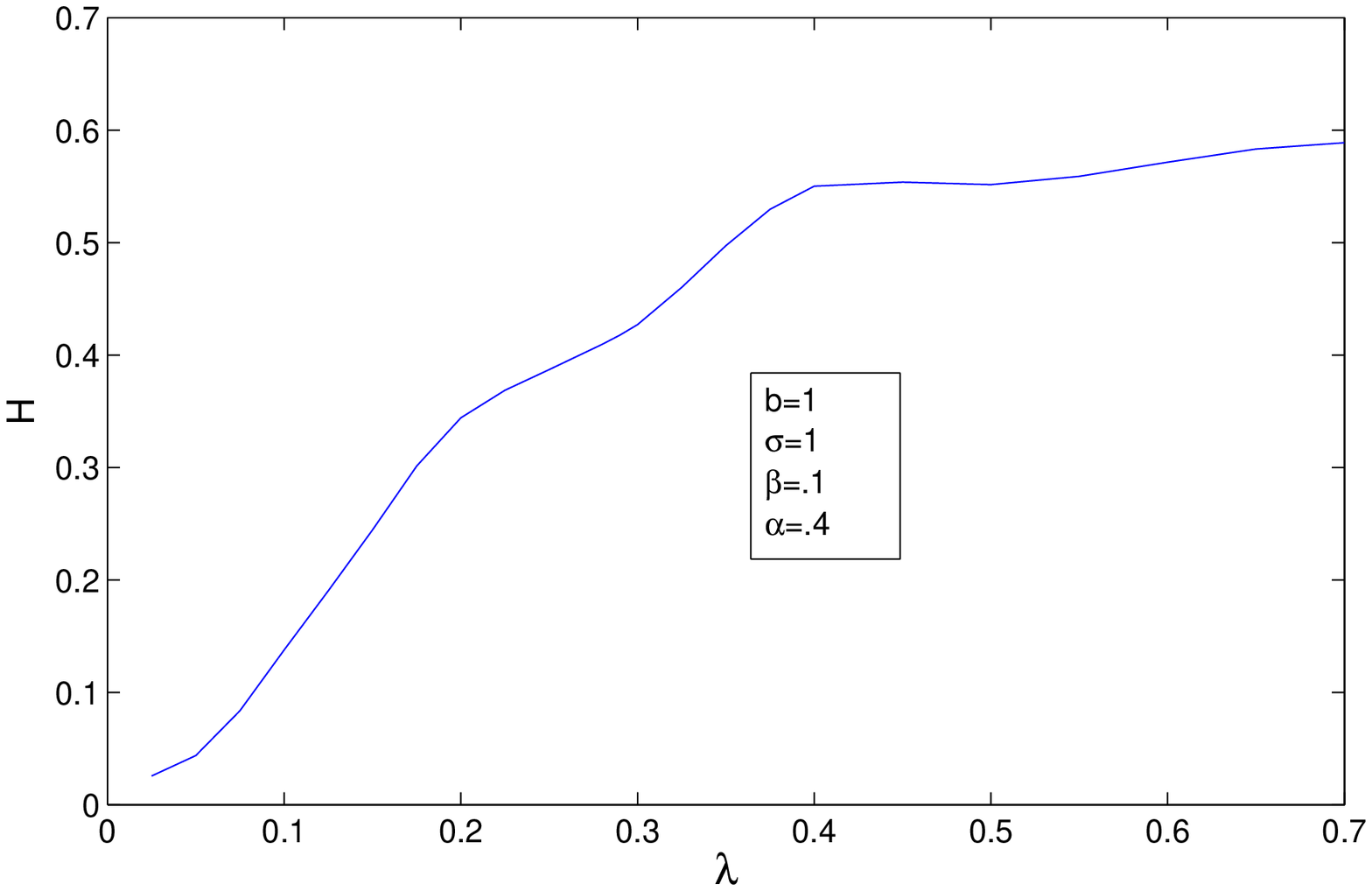}
\caption{(color online). Mutual information versus lambda
($\lambda=\Lambda /\Lambda_p$) for SPDC. ($\alpha$, $\beta$, and
$\sigma$ are in units of $L^{-1}\omega_0^{-1}$, $L^{-1}
\omega_0^{-2}$, and $\omega_0^{-2}$, respectively. $b$ is in units
of $\omega_0^{-2}$, while $\lambda$ is dimensionless.)}
\label{lambdafigx}
\end{figure}

\begin{figure}
\centering
\includegraphics[totalheight=2in]{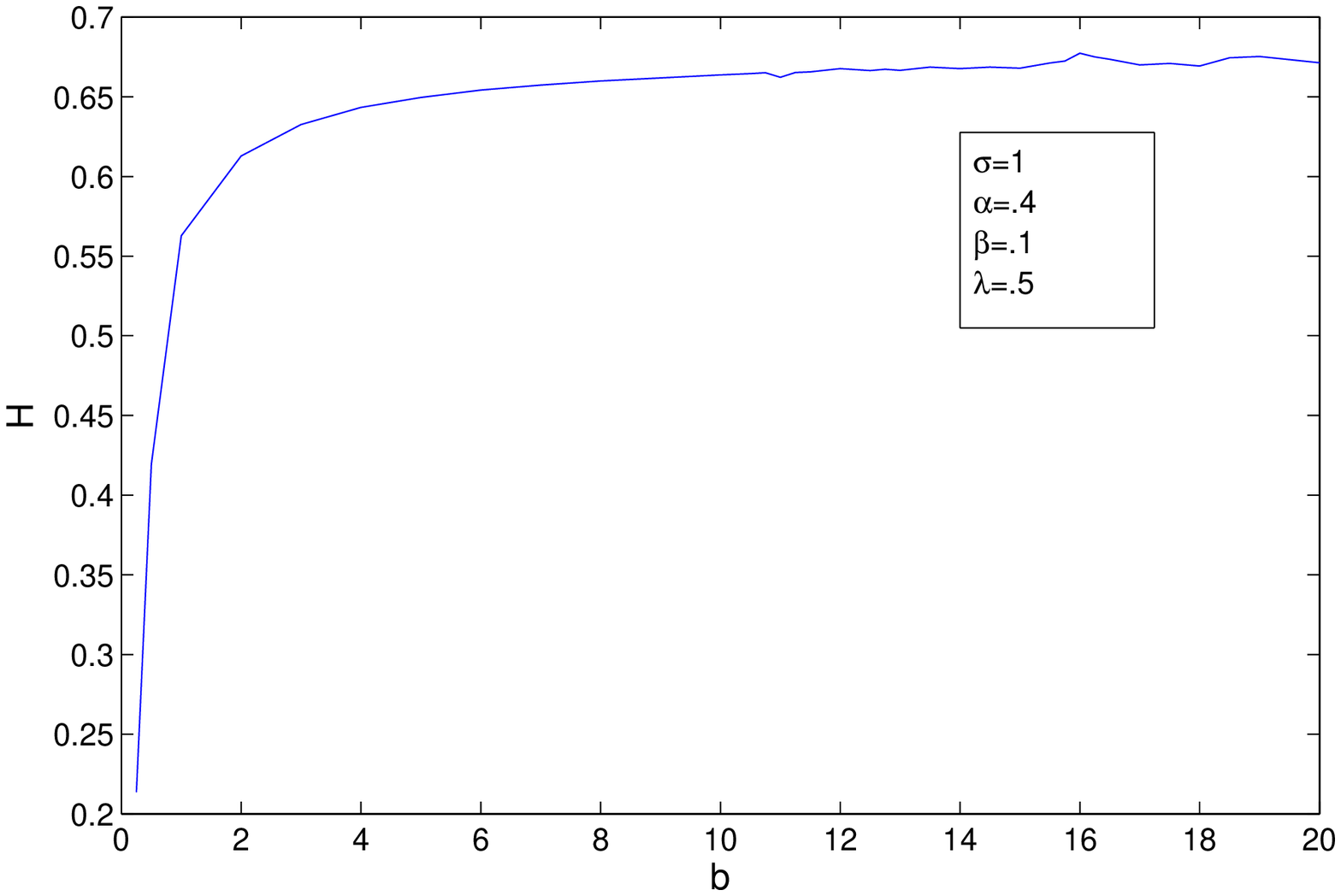}
\caption{(color online). Mutual information versus b
($b={{L_c}\over 2}\Lambda_p $) for SPDC. ($\alpha$, $\beta$, and
$\sigma$ are in units of $L^{-1}\omega_0^{-1}$, $L^{-1}
\omega_0^{-2}$, and $\omega_0^{-2}$, respectively. $b$ is in units
of $\omega_0^{-2}$, while $\lambda$ is dimensionless.)}
\label{bfigx}
\end{figure}

Given these output probabilities, the mutual information can once
again be computed. In contrast to the previous sections, the
analytic forms of the probabilities are too complicated to be
enlightening, so we proceed to numerical calculations.  Some
examples are graphed in figures~\ref{sigmafigx} to~\ref{bfigx}.
The plots are expressed in terms of the new parameters $b
=\Lambda_p L_c$ and $\lambda={\Lambda \over {\Lambda_p}}$.

Examples of the dependence of $H$ on the parameters of the pump
beam ($\sigma$), interferometer ($\alpha$, $\beta$), and nonlinear
crystal ($\lambda$, $b$) are given in figures~\ref{sigmafigx}
through~\ref{bfigx}. We see that, although $H$ decays overall with
increasing values of the dispersion paramaters in the
interferometer, $\alpha$ and $\beta$, there are oscillations
superimposed on the decay, which are especially noticeable at low
values of $\alpha$ and $\beta$. This effect was in fact also
present in the simplified SPDC model of the previous sections, but
in the latter case the oscillations were too weak to be visible on
the graphs. We see also that as either $b$ or $\lambda$ increases
(or equivalently, as $\Lambda$ or $\Lambda_p$ increases), the
plots approach those of the simplified SPDC model. Since $\Lambda$
and $b$ are proportional to the crystal length, this means that
the simplified SPDC model is an increasingly better approximation
to real SPDC for longer crystals. It also appears from the
numerical simulations that for a given set of parameter values
$\alpha$, $\beta$, $L$, and $\sigma$, the simplified SPDC model
provides an upper bound to $H$ for the real SPDC cases with the
same parameter values. The maximum information content clearly
occurs for low dispersion in the interferometer, long nonlinear
crystals, and large mismatch at $\omega_0$ between signal and
idler inverse group velocities in the crystal (large $\Lambda$).

\section{Conclusions}\label{conc} In this paper, we have examined the effect of
dispersion on the mutual information that interferometric
photon-detection measurements carry about phase shifts. We have
looked at a number of different situations involving two
interferometer set-ups and several different types of
non-classical input states. Comparing the results, we now have a
precise and quantitative means to measure the relative merits of
different input states for various input-parameter ranges. As a
by-product, we have shown that in some circumstances, parametric
downconversion can be approximated by a much simpler model that is
amenable to exact analytical analysis.

Returning to the original question of which input state yields the
most information about the phase shift, the graphs of the previous
sections yield fairly clear results. Restricting discussion to
MZ interferometers for simplicity, we can see that for
quantum interferometry in the presence of dispersion the entangled
photon pair produced by downconversion has a clear
advantage over other cases when input to a single port (Fock
state input). This advantage does not exist in the case of an
dispersionless interferometer, in which case the
presence or absence of frequency correlations becomes irrelevant
for the information content. The only situation we have found in which another input is superior to the
frequency-anticorrelated Fock input is when $\alpha$ is large but
$\beta$ small, in which case the anticorrelated {\it dual} Fock
input is superior. These conclusions all hold when the simplified
downconversion model of section~\ref{spdc} is a good
approximation; the results of section~\ref{spdcreal} imply that
such conclusions weaken as the crystal
becomes shorter.

\begin{acknowledgments} This work was supported by a U. S. Army Research Office (ARO)
Multidisciplinary University Research Initiative (MURI) Grant; by the Bernard M. Gordon
Center for Subsurface Sensing and Imaging Systems (CenSSIS), an NSF Engineering Research
Center; by the Intelligence Advanced Research Projects Activity (IARPA) and ARO through
Grant No. W911NF-07-1-0629.
\end{acknowledgments}

\end{document}